# Diffuse Matter and Cosmogony of Stellar Systems in Researches by G.A. Shajn

*N. I. Bondar'*

Crimean Astrophysical Observatory RAS, Nauchny, Republic of Crimea, 298409, Russia; otbn@mail.ru

## Abstract

The main topic of long-term researches by G.A. Shajn is the nature of diffuse matter, its distribution in the Galaxy and extragalactic systems, interaction with the interstellar medium and hot stars, the formation of emission and reflection nebulae and stars have become the main theme of researches by Shajn during some years. Based on the analysis of experimental data, mainly photographic observations of nebulae in the Milky Way and extragalactic systems, he made conclusions and suggested well-founded hypotheses on a wide range of the considered problems, including those related to cosmogony.

The structure of nebulae, their masses and sizes give a reason to the conclusion that most of them are formed not in the process of ejection of matter from the stars, but this is the objects, which are born and evolved, and quite often comprised in giant conglomerates of gas, dust and stars. The distribution of O-stars and nebulae in spiral branches points to their genetic relation and the fundamental role of the interstellar medium as a source of their formation. The structural features of nebulae are determined by the action of magnetohydrodynamic forces. Magnetic fields in a galaxy control the motion of diffuse gas-dust matter and ensure the maintenance of its spiral structure. These ideas continue being developed in modern directions of astrophysics.

**Key words:** *galaxy—interstellar matter; diffuse nebulae; cosmogony*

## 1 INTRODUCTION

An intensive study of interstellar matter, its properties and distribution in our Galaxy and extragalactic systems was began in the middle of the 20th century and posed a number of problems concerning the origin of stars and stellar systems, their stability and evolution (Ambartsumian 1933, 1949; Whipple 1946; Bok & Reilly 1947; Sharpless 1954). The first ideas and theories proposed in a number of works were based on the results from scanty observational data (Weizs.acker 1951; Blaauw 1952; Sharpless & Oster-brock 1952; Oort 1954; Oort & Spitzer 1955).More extensive facts obtained by different methods had required for their verification and development.

Long-term investigations of nebulae in the Milky Way were carried out at Crimean Astrophysical Observatory (CrAO) for about 20 years according to the program developed by G.A. Shajn.

Grigory Abramovich Shajn (1892–1956) is a famous astrophysicist, a member of the Academy of Sciences of the USSR and a number of astronomical societies. His scientific path began at Pulkovo Astronomical Observatory in 1921, in 1925 he was sent to Simeiz Observatory, which under his leadership became the leading astrophysical observational base of the Soviet Union. Investigations of small planets and comets, the Sun, variable stars and galaxies were carried out there. Shajn was the initiator of the foundation of Crimean Astrophysical Observatory and its Director in 1944–1952. His bibliography contains about 150 works concerning different astronomical objects - meteors, comets, small planets, stars, nebulae, and galaxies. His results on the rotation of stars, the features of spectra of long-period variables and the abundance of carbon isotopes in cold stars, on the study of nebulae in the Milky Way and the interstellar medium made contribution to the physics of stars and galaxies and are determined the development of a number of directions in this field. Shajn's biography has been presented in articles by Pikel'ner (1957), Struve (1958), Chuvaev (1995).His scientific heritage was generalized in reviews of different years (Pikel'ner 1957; Struve 1958; Pronik 1998, 2005, 2008; Pronik & Sharipova 2003) and in proceedings of the memorial conference published in *Izv. Krym. Astrofiz. Obs. (BCrAO) v. 90, 1995.*

In the late 1930s Shajn studied the distribution of matter in the Galaxy. He obtained an unexpected result which was named the "Shajn Paradox". Observations have shown the absence of correlation between the reddening of stars and the brightness of the Milky Way. Stars with a small reddening were detected in dark areas, and in contrary, stars with a strong color excess were found in light areas (Shajn 1937). This fact found an explanation in the cloudy structure of the Milky Way, whose apparent brightness is associated with near clouds, at a distance of 100–150 pc from the Sun. Also Shajn & Dobronravin (1939) carried out the study of the continuous spectrum of the Milky Way.

In the years of the World War II the 1 m telescope at Simeiz Observatory was demounted and after the war the spectral investigations could not be continued. Shajn decided to choose a direction of research in which with small instruments it could be possible to obtain actual results relevant to the physics of interstellar medium and to cosmogony of stellar systems (Shajn & Gaze 1951). He saw the development of questions of cosmogony in the solution of problems relating to the physical nature of nebulae, the diversity of their physical characteristics, the distribution in the galaxy, and the interaction of interstellar matter and stars. Later, the important part of this program became a study of magnetic fields and their role in maintaining the structure of nebulae and spiral arms of galaxies.

The systematic survey of nebulae was started by G.A. Shajn in 1949, together with V.F. Gaze, and within two years they obtained images of nebulae in the Milky Way in the belt of latitude limited to $\pm 10^\circ$ (Shajn & Gaze 1951). At the longitude the observations covered the Milky Way from Sagittarius to Monoceros. The researches have been based on the experimental data obtained by Shajn together with colleagues in the Crimean Astrophysical Observatory (CrAO). Shajn used high-speed cameras (F:1.4) with a diameter of 450 mm and 640 mm for photography and narrow-band filters that extracted in a spectrum strong emission lines and narrow bands in the continuum. This method of observations and the appropriate choice of exposure made it possible to obtain unique images of nebulae with many details invisible in early images of already known nebulae and discover new nebulae.

The results obtained during the first two years Shajn presented in his report at the VIII General Assembly of IAU in 1952 in Rome (Shajn & Gaze 1954a). These studies were recognized important, and it was decided to create a new catalog of gas nebulae. For this aim the Sub-Commission of IAU was organized and V.F. Gaze actively participated in it until 1953.

Vera Fedorovna Gaze (or V.Th. Hase) (1899–1954) was a researcher at Crimean Astrophysical Observatory, the author of about 40 papers in the field of minor planets, stellar spectroscopy, and over her last years she studied the diffuse nebulae and the structure of our Galaxy (Shajn 1955a). In 1955 Shajn and Gaze published a catalogue (Gaze & Shajn 1955), compiled on the basis of four lists of nebulae prepared in 1950–1954. The new catalogue included of 286 nebulae, objects discovered in Simeiz Observatory are designated in it by the letter "S" (Simeiz). This catalogue and the published one by Sharpless (1953) provide complete information about the location of nebulae near the galactic plane. Photos of 48 bright nebulae were published in the Atlas of diffuse nebulae (Shajn & Gaze 1952a). To solve some tasks, observations of several nebulae were carried out with the nebular spectrograph at Simeiz Observatory (Pikel'ner 1954), and polaroids (Shajn et al. 1955b), the results in the radio range were used also (Shklovsky & Shajn 1955).

Wide-field photographic observations and observations with an objective prism were carried out since 1950 in the specified band of the Milky Way in 13 selected areas 10°x10° in size to study absorption at different distances, determine the connection between nebulae and hot stars, and also to identify regions of star formation. These works were continued until 1967, reviews of the results are presented by Pronik (1998, 2005) and Pronik & Sharipova (2003). The obtained photographic material is preserved in the archives of CrAO (Bondar 1999; Shlyapnikov et al. 2015; Shlyapnikov et al. 2017), at the International Center in Sofia (Bulgaria) and in a digital version in the Virtual Observatory of CrAO (*http://www.craocrimea.ru/~aas/Projects/SPPOSS.html*).

Investigations of diffuse nebulae were performed by Shajn in the last years of his life and he could not summarize them into a monograph. He presented some of the results in the report which prepared to IAU Symposium No.6 (Shajn 1958b). Besides the above catalogue and the Atlas of diffuse nebulae he has published more than 40 articles in which he extensively analyzed the images of nebulae and dis-cussed the hypotheses and conclusions about the nature and de-velopment of nebulae and galactic systems. This brief review contains the main results and con-clusions published by Shajn in his papers and in a series of works with co-authors on the follow-ing topics: the structure of diffuse nebulae, their physical parameters, the relationship with OB-type stars, the distribution of nebulae in the galaxy, the magnetic field, and the interstellar medi-um.

## 2 STRUCTURE OF NEBULAE

The structural features of nebulae were studied from the images obtained in the hydrogen line $H_\alpha$ with different exposures and ~ 4′ in the scale. At this the distribution of matter, its concentration in individual light and dark details, position relative to ionizing stars were taken into account. In addition to the known types of nebulae, Shajn and Gaze have allocated two new types. One of them were named filamentary nebulae. This type includes nebulae similar to the nebula Simeiz 147 in Aurigae discovered by Shajn and Gaze in 1951 (Shajn & Gaze 1952e; Shajn 1955d). The nebulae in which gaseous matter is concentrated at a distance from the center were named pe-ripheral (Shajn & Gaze 1953b). Extensive low-luminosity nebulae known as hydrogen fields were also identified (Shajn & Gase 1952b,1954a).

Peripheral nebulae are different in structure and look like an extended gaseous clouds, arcs or rings at some distance from one or a group of O−B0 stars located in the central region of the nebula. They are similar to planetary nebulae in this, but brighter and more massive. The list prepared by Shajn & Gase (1953d) includes 20 nebulae of this type in our galaxy and 8 in extra-galactic systems. Taking into account that only the brightest nebulae were considered, they have made conclusion that this type of nebulae is rather numerous in our and other galaxies.

One more their new result is that nebulae are not amorphous and stationary objects. Obvious de-tails in nebulae − rings, stripes, segments, streams, filaments, branches − point to not only morphological differences (Shajn 1958a). They show that nebulae are the objects with a complex dynamics manifesting turbulent motion, collisions and interactions of individual fragments and layers. The characteristic feature of different types of nebulae, including planetary and reflection nebulae (Pleiades, for example), is filaments and layers. Filaments in nebulae have different shapes, sizes and brightness, they may be found both on an edge and inside the nebula, in some nebulae they have a preferential direction. The outward motion of matter in nebulae leads to the dissipation of matter and formation of low-density and low-luminosity clouds around them. The scattering of matter causes uneven edges of the nebula, blurring of boundaries, arising of low luminosity matter around the main nebula and even the appearance of the separate parts.

Shajn revealed that the characteristic feature of nebulae is uneven distribution of brightness in them, except for hydrogen fields. This unevenness is caused by their structure, stratification, and by the presence of very small dark nebulae, globules and more significant dark formations which are closely associated with related to the origin and evolution of the nebula, its physical state, dynamics, interaction with the interstellar medium. The nebulae include gaseous and solid parti-cles, however, the number of dust particles is extremely small in gaseous nebulae (Shajn et al. 1954, 1955a). Shajn was the first to find the relative abundance of dust in the diffuse nebulae.

## 3 PARAMETERS OF NEBULAE

To study the origin of nebulae and to solve the question about the relation with the embedded stars Shajn determined sizes, masses and densities of nebulae. To estimate mass and density, it is necessary to know linear sizes and shape of the nebula, its surface brightness from 1 square arc second. Calculations of these parameters were made for several bright nebulae, their surface brightness in the $H_\alpha$ line was converted into absolute values, in erg cm$^{-2}$ s$^{-1}$ (Shajn & Gase 1952d). In addition to nebulae in the Galaxy, he studied extragalactic nebulae M33, M31, M101, NGC 6822 (Shajn & Gase 1953e) and found that the most nebulae have masses of Mn < 10M$_\odot$. Small nebulae are comparable in mass to the Sun or less, filaments are the formations with masses of 0.01M$_\odot$. Massive nebulae have Mn > $10^3$M$_\odot$, they are not numerous in our Galaxy. Giant massive gaseous or gas-dust clouds with masses of $10^4 - 10^5$M$_\odot$ were detected in extragalactic nebulae. Masses of nebulae have been determined with an accuracy of 15–30%. These are summarized errors, involving errors of the method, features of the structure of nebula and also errors in determining sizes of nebulae and distances to them. However, even with large errors, one can confirm confidently that the mass of nebulas matter exceeds the mass of hot stars in it.

The sizes of nebulae also are in wide ranges – from compact nebulae with radii in fractions of a parsec to giant clouds with R = 8–60 pc. Filaments exceed a length of 1–4 pc, their width and thickness are fractions of a parsec. The density of protons in the nebula is higher than in the interstellar medium – from 10 to 100 and more per 1 cm3, and in filaments it becomes very high exceeding $10^3$ protons per 1 cm$^3$.

Thus, the obtained results show that the ejection of matter from stars or their destruction lead to the appearance of some types of low-mass nebulae, but these processes cannot produce the formation of massive nebulae.

## 4 CONNECTION WITH HOTWR- AND OB-STARS

Over three years of investigations, Shajn & Gaze (1953a) obtained images of 300 nebulae and performed the statistical analysis of this material and 536 Wolf-Rayet (WR) and of O–B1 stars to determine the correspondence in location of nebulae and stars and to consider the question about their random or genetic connection. Hot stars in the investigated galactic belt turned out to be located mainly in regions of nebulae localization, but maxima of their concentration do not always coincide. In some areas there are deviations from this relationship, for example, in Perseus the region with a white supergiant cluster has no gaseous nebulae.

The brightest and isolated nebulae are related to hot stars, this connection increases with increasing temperature of the star – from B1- (19%) to O- and WR-stars (56% out of the considered stars of these types). Among bright, massive nebulae, about 85% are associated not with one, but with a group of O-stars and multiple systems. Stars of later spectral types – B2–B5 stars – are associated with gas-dust and dust nebulae (Shajn et al. 1955a). The diffuse nebulae with a central position of O–B1 stars (peripheral, compact spherical) are genetically related to hot stars responsible for their illumination.

The question about a random or genetic relation is solved ambiguously in a case if an ionizing star is at the edge of the nebula, if the nebula is extensive, but the position or number of stars is insufficient for its illumination, if there are no hot stars in the nebula. In these cases, other mechanisms of luminosity are considered, and the dominant role in the interaction with stars passes to diffuse matter, which mass exceeds the mass of the embedded stars. It should be noted that the extended up to several dozen parsecs hydrogen fields with a low electron density are very weak to be studied by the used methods. Some of them are definitely associated with O–B0 stars, but in some cases these stars could not be detected simply.

## 5 DISTRIBUTION OF NEBULAE IN THE GALAXY

The discovery of several hundred nebulae at penetrating into the depth of the Galaxy up 2–3 kpc means that gas and gas-dust nebulae are numerous objects of the Galaxy and extragalactic systems (Shajn 1958a). The distribution of nebulae is not random. The groups of nebulae, probably related genetically, are concentrated in the spiral arms of galaxies almost along all their length. A group involves several nebulae located approximately at the same distance from us. Shajn identified 21 groups of nebulae in our Galaxy (Shajn & Gaze 1953d; Shajn 1954, 1956a), the largest of them are in Orion, Sagittarius and Cygnus, their extension reaches several hundred parsecs. At the same time, there are fields with a small number of nebulae or their absences, for example, in the regions of constellations Aquila and Perseus (Shajn & Gaze 1953a, 1954a).

In the spiral arms a non-uniform distribution of nebulae is observed with significant fluctuations in concentration of gas-dust matter on the scale of 150–300 pc and more. Along with the bright emission nebulae, which were found in nearby extragalactic objects, there exist many nebulae of medium and small scale. Shajn made a conclusion that the characteristic feature of spiral galaxies is the presence of a large number of gaseous nebulae. However the spiral arms include conglomerates of gas, dust and stars. The dust is found in both reflection and emission nebulae, including planetary ones. A dark nebula can penetrate or surround a diffuse nebula, and dust can be concentrated in the form of filaments and globules and other dark structures (Shajn & Gaze 1953a; Shajn et al. 1954, 1955a).

## 6 MAGNETIC FIELDS IN THE INTERSTELLAR MEDIUM AND NEBULAE

In searching for understanding inner processes in the nebulae Shajn (Shajn & Gaze 1954a,b) came to the idea that difficulties in explaining structural features of nebulae, their elongation and preferential orientation only by action of hydrodynamic processes, tidal interactions and differential rotation of the Galaxy, can be overcome if to assume the hypothesis on interaction of hydrodynamic and electromagnetic phenomena. The discovery of polarization of stellar light in 1949 by Hiltner (1949), Hall (1949) and Dombrowskiy (1949) has been the observational fact about the presence of the magnetic field of an order of $10^{-5}$ gauss in the interstellar medium. Taking into account that gaseous and dust nebulae are an electrically conductive medium in which non-stationary motions occur, Shajn & Gaze (1951) made a working assumption that there are inner magnetic fields in the nebulae.

In series of papers in 1952–1955 Shajn considered arguments to confirm and develop the offered hypothesis. He paid attention that dark and some emission nebulae are elongated of over distances up to several parsecs and the orientation of nebulae and filaments in them is the same (Shajn & Gaze 1952c). The certain direction of elongation is observed not only in the some nebula but it is characteristic feature in some wide region. Comparison of the observed elongation with polarization data in this region has shown that nebulae are elongated in the direction coincident with the plane of polarization (Shajn 1955c). Basing on these facts Shajn made a conclusion that magnetic field influences the nebula matter, having the ability to dissipate and move outward, decelerates it in the direction perpendicular to the lines of force and allows moving along them.

According to calculations the velocity of macroscopic motions of 1–2 km s$^{-1}$ in dust nebulae and 10–15 km s$^{-1}$ in an emission ones, the external magnetic field is able to stretch a nebula up to 20–30 pc during $10^7$ and $10^6$ years, respectively. In diffuse nebulae the elongation of globules and the motion of matter between them are also explained by the influence of magnetic fields. The scattered matter is oriented along the lines of force, even if the nebula is decaying. As it follows from polarization measurements at the low galactic latitudes ($< 10°$) the elongation of nebulae as well as filaments in them is almost parallel to the galactic equator and coincides with the direction of the external regular magnetic field. In higher latitudes the nebulae are elongated at a significant angle to the galactic equator, what points to the action of the local magnetic field in the region of their location (Shajn 1956d).

Shajn has shown that magnetic field in the interstellar medium, dynamic and other physical processes in it are in connection (Shajn 1955c, 1956b,c; 1957). Turbulent motions inside the nebula lead to an increase of the inner magnetic field, at the same time this field is inhomogeneous due to fluctuations in density. Magnetic field lines frozen into the nebula matter thicken in high-density regions. The presence of the irregular magnetic field (up to $10^{-4}$ gauss) allows one to interpret the concentration of matter and the brightness increase in the central part of some nebulae (the Crab Nebula, IC 1805, NGC 1976 in the Orion Nebula).

Shajn's assumptions about the existence of magnetic fields in nebulae and galaxies found independent evidences in works by Pikel'ner and Shklovsky. Pikel'ner (1953) showed that the magnetic field maintains cosmic rays in the halo and disk. Shklovsky (1953) explained the radiation of the Crab Nebula in radio and optical spectrum by emission of relativistic electrons in the magnetic field. He discussed this mechanism within the problem of cosmic ray generation during an outburst of supernovae and novae. In the 1950s, the Crab Nebula is a puzzle object, a remnant of supernova 1054 discovered by Chinese astronomers, and it was studied intensively by different methods. Measurements of the polarization in different parts of the Crab Nebula (Shajn et al. 1955b) have shown that the magnetic field in the amorphous central part is more regular than at the periphery. The polarization in the nebula is caused not by absorption inside or outside the nebula, but due to the magnetic bremsstrahlung radiation of relativistic electrons, that produces both continuous emission and polarization The role of magnetic field and cosmic rays in generation of optical and radio emission of the Crab Nebula and in the secular acceleration of its filaments was shown by Pikel'ner (1956, 1964) within magnetohydrodynamics developed in his theoretical works. In the frame of magnetohydrodynamics the structure of nebulae associated with supernovae, the orientation of arcs and filaments in them exhibit a pattern of distribution of the magnetic field on the star. This field affected the initial conditions of the ejecting stellar envelope and to some extent determined the currently observed distribution of the gas matter.

However, as noticed by Shajn (1955b, 1958b), the inner magnetic field, dominated in some nebulae, does not provide their stability. This follows from the fact that fragmentation, dissipation and destruction are typical for many of them. The interstellar magnetic field has a significant effect on the distribution of diffuse matter along the spiral arms of the Galaxy and, preventing the scattering of gas-dust clouds, maintains the stability of the structure of the Galaxy and extragalactic systems. As follows from a study of the magnetic field in the solar vicinity the direction of the galactic regular magnetic field is subject by fluctuations on the order of 1000 pc (Shajn 1957).

Shajn's conclusions about magnetohydrodynamics processes in diffuse nebulae and interstellar medium have been confirmed and developed in theoretical works by Pikel'ner (1961, 1966, 1968, 1970), Kaplan & Pikelner (1970, 1974 ) and others.

Solomon Borisovich Pikel'ner (Pikelner) (1921–1975) began to study the nature of nebulae and physical processes in the interstellar medium together with Shajn. He considered the problems of radiation of nebulae, turbulent motion, influence of different kinds of instabilities in the medium and in the magnetic field on radiation and features of the structure of nebulae, in particular, in the Crab nebula. He created a new branch in astrophysics – physics of cosmic plasma, in which physical processes on different astrophysical objects and in the interstellar medium are studied within the influence of the magnetic field.

# 7 COSMOGONY OF DIFFUSE AND DUST NEBULAE

In the relationship between stars and nebulae, a different nature of their relationship may be traced, depending on the position of stars and nebula mass. The matter ejected by the star can be a cause for the formation of only certain types of low-mass nebulae. In bright massive nebulae stars or groups of hot O–B stars located in the central region of nebulae and providing their illu-mination are related genetically. But Shajn notes that this relationship, followed from the statistics of distribution of stars and nebulae, is more deep and complex (Shajn & Gaze 1953a; Shajn 1955b). There is a reason to suppose that in the region where hot stars and their groups have been formed, simultaneously the gas nebulae arise and hence, nebulae and stars are formed in the same epoch, perhaps from one source, and then they evolve by their own ways.

In evolution of nebulae it should be taken into account the interaction of solid particles and gas, i.e. the initial density of the medium and consequent concentration of particles, which varies in time. During $10^6 - 10^7$ years the dust component in the gaseous nebula increases and this nebula becomes a gas-dust nebula (Shajn et al. 1954a, 1955a). The offered qualitative explanation is one of the possible or probable, but considering this problem it becomes obvious that nebulae are evolving objects participated together with stars in the exchange of matter with the surrounding interstellar medium, which plays a dominant role in the formation both nebulae and stars.

From the point of view of the fundamental contribution of diffuse matter, one should consider the dynamics and evolution of giant diffuse nebulae, where complex interactions of stars, ionized gas and dust should occur. In such systems objects of various ages exist simultaneously, including O-associations and the youngest multiple stars of the Trapezium type (Shajn 1955e). Diffuse matter and associated bright stars are concentrated in the spiral arms of galaxies (Shajn & Gaze 1953b). In the cosmogonic aspect, the entire gas population of the spiral branch can be considered as a whole, as a system of higher order. At the same time fluctuations of different scales are observed in the spiral branches, there may be distinguished the extended systems of nebulae, regions with small concentrations or with total absence of them. Systems of nebulae refer to the young objects, they are detected along the entire spiral, thus the spiral branch in whole presents the young formation (Shajn & Gaze 1954c; Shajn 1958a,b).

The magnetic field contributes to the elongation of nebula matter and its motion along the spiral arms, preventing scattering. The spiral branch can be regarded as a directed cosmic gas flow, with thickenings and knots, where the genetically related groups of nebulae are concentrated. Such groups are very extensive reaching hundreds of parsecs and more (Shajn & Gaze 1952c, 1953c, 1954a; Shajn 1955c,d).

Based on the age of diffuse nebulae and the spatial velocity of 10 km s$^{-1}$, it follows that they were born not far from the place where they are currently, i.e. at a distance of several hundred parsecs. But since the gas nebulae are visible along the entire spiral branch, it means that they originated along the entire arm simultaneously. In any case, it cannot be assumed that they were formed in some central region and then spread along a spiral. Shajn noted that only the first steps had been made in understanding the formation of gas-stellar complexes and for the complete investigation of this problem the study of neutral hydrogen clouds in galaxies must be involved. During the time of existence of nebulae as a result of their scattering along the spiral branches, it might be expected the accumulation of the large masses of neutral hydrogen.

Significant masses of it were detected from radio observations at the wavelength of 21 cm. But this does not mean that neutral hydrogen is the final product. The initial conditions for the formation of nebulae and stars may be created in cool clouds of a neutral gas ( Shajn 1954, 1958b; Shajn & Gaze 1954c).

## 8 CONCLUSIONS

The high-quality images of the known and first detected nebulae in the belt of the Milky Way derived by G.A.Shajn and V.F. Gaze have exhibited the structural diversity of these objects and allowed one to put forward a number of hypotheses to explain their nature and origin. In works by Shajn it has been shown for the first time that nebulae are not amorphous formations, but have a certain structure and evolve on the time scale of 107 years. A characteristic feature of nebulae is the presence of filaments oriented in a certain direction determined by the magnetic field in the given region. In the problem of the origin of stars and nebulae Shajn believed that it cannot be excluded the fact that some low-mass nebulae have arisen due to ejections from stars or supernovae burst, but in general the interstellar medium and the conditions in it have a dominant role in the formation of both objects. This conclusion is based on the obtained by Shajn results that masses of many nebulae are larger than the total masses of ionizing stars. The comprehensive study of images of nebulae leads Shajn to the discovery that in similar to the stars in O-associations, emission nebulae were born in groups and in the same places as O-associations, what indicates their genetic connection. Based on this, he came to the conclusion that stars and nebulae have been arisen in the same region and possibly from one source. They should be studied as objects which belong to one system, especially at the early stage of development. This was a new approach to the question of whether stars originate from nebulae or vice versa.

Shajn found that diffuse nebulae contain dust and he received the first estimation of its relative content. He established that in the emission and reflection nebulae there is a connection between the gaseous and dust matter. This is caused by the motion of nebulae relative to each other and in the interstellar medium, as well as by physical processes - condensation and evaporation, which lead to the formation of dust grains from gaseous particles and vice versa. The interstellar medium, enriched with stellar matter and scattered diffuse matter, is a material for the formation of stars and nebulae. In the absence of such cyclical exchange during the time of existence of the Galaxy, the amount of the scattered matter should be 100 times larger than the currently observed concentration in diffuse nebulae.

Methods applied by Shajn to study nebulae made it possible to carry out not only qualitative analysis, but also to obtain quantitative estimates of parameters of nebulae, to estimate the distance to them and to represent their distribution as the objects which belong to spiral arms of the galaxy.

Shajn came to the conclusion that when studying nebulae, the interstellar medium and the structure of the galaxy, it is necessary to take into account the existence of magnetic fields, internal and external, i.e. on the basis of magnetohydrodynamic processes. The magnetic field provides retention of the scattered matter in the region of spiral arms and thus maintains the stability of the structure of our Galaxy and spiral and irregular extragalactic systems.

In addition to the results presented here, Shajn considered questions about the nature of luminosity of different types of nebulae, the formation of HII and HI fields, and sources of radio emission. The tasks formulated by Shajn and developed in his methods have proved to be productive in studying questions on cosmogony. The whole complex of his works on this topic had a significant impact on the subsequent works of Pikel'ner, Kaplan and other astrophysicists, founders of the physics of the interstellar medium, cosmic gas dynamics, and extragalactic astronomy.

## Acknowledgements

The author is grateful to the referee for useful remarks and recommendations. Many thanks to R.E. Gershberg and A.A. Shlyapnikov for discussing the paper, and Z.A. Taloverova for technical support.

The work was partially supported by the Russian Foundation for Basic Research and the Ministry of Education and Science of the Republic of Crimea, the project 16-42-910595 r_a.